\documentclass[a4paper,11pt]{article}
\usepackage{jinstpub} % for details on the use of the package, please
\usepackage{epsfig}
\usepackage{epstopdf}
\usepackage{graphicx}
\usepackage{subfigure}
\usepackage{hyperref}
\title{\boldmath Further studies of proportional electroluminescence in two-phase argon}

\author[a,b]{A.~Bondar,}
\author[a,b]{A.~Buzulutskov,}
\author[b]{A.~Dolgov,}
\author[a,b]{E.~Frolov,}
\author[a,b]{V.~Nosov,}
\author[a,b]{V.~Oleynikov,}
\author[a,b]{L.~Shekhtman,}
\author[a,b,1]{E.~Shemyakina\note{Corresponding author.}}
\author[a,b]{and A.~Sokolov}

\affiliation[a]{Budker Institute of Nuclear Physics, \\Lavrentiev ave. 11, Novosibirsk 630090, Russia}
\affiliation[b]{Novosibirsk State University,\\ Pirogova st. 2, Novosibirsk 630090, Russia}

\emailAdd{E.O.Shemyakina@inp.nsk.su}

\abstract{The study of proportional electroluminescence in two-phase argon is relevant in the field of noble-gas liquid detectors for dark matter search and low-energy neutrino experiments. In this work, we continued to study proportional electroluminescence (EL) in two-phase argon doped with a minor (9 ppm) admixture of nitrogen, in the VUV, UV and visible spectral ranges.  We confirmed the effect of enhancement of the EL yield, as well as the presence of non-VUV component in addition to that of VUV, in proportional electroluminescence in two-phase Ar. On the other hand, the contribution of non-VUV component determined here within the model of N$_2$ emission in the UV, turned out to be insufficient to explain the EL yield enhancement effect. Accordingly, the problem of proportional electroluminescence in two-phase Ar remains unresolved.}


\begin{document}
\maketitle
\flushbottom

\section{Introduction}

The study of proportional electroluminescence in two-phase argon is relevant in the field of two-phase detectors for dark matter search \cite{DarkSide15} and coherent neutrino-nucleus scattering \cite{CoNu04,CoNu09} experiments. In two-phase detectors, the S2 signal (induced by the primary ionization in the noble-gas liquid), is detected through the effect of proportional electroluminescence (or secondary scintillation) in electroluminescence (EL) gap located directly above the liquid-gas interface \cite{NobleRev13}. In proportional electroluminescence the energy provided to the electrons by the electric field is almost fully expended in atomic excitations producing Ar$^{\ast}(3p^54s^1)$ states. These are followed by the photon emission in the Vacuum Ultraviolet (VUV), around 128 nm, due to excimer (Ar$^{\ast}_2$) productions in three-body collisions and their subsequent decays (see \cite{ArXeN2Proc17} and references therein).

In the presence of nitrogen admixture to gaseous argon the electroluminescence mechanism might be modified. Namely, the excimer production (and hence the VUV emission) can be taken over by that of excited N$_2$ molecules (due to excitation transfer from Ar to N$_2$ species), followed by their de-excitations in the near Ultraviolet (UV) through the emission of the so-called Second Positive System (2PS), at 310-430 nm \cite{ArXeN2Proc17}. Such a VUV-to-UV conversion, taking place directly in the detector medium, would substantially enhance the photon collection efficiency for the S2 signal due to avoiding problems with re-emission and total reflection losses in the Wavelength Shifter (WLS) film, which would be used otherwise.

It is known that in gaseous argon at room temperature such a VUV-to-UV conversion of electroluminescence can be effectively performed by doping with nitrogen at a relatively large content, of 0.2-2\% \cite{Policarpo67,Takahashi83,Kazkaz10}. Recently proportional electroluminescence of gaseous Ar in the VUV has for the first time been studied at cryogenic temperatures, in the two-phase mode \cite{CRADPropEL15,CRADELGap17}: the EL yield was measured in Ar, with a minor ($\leq$50 ppm) admixture of N$_2$ that might be typical for large-scale liquid Ar experiments.   The EL yield was found to be substantially enhanced, by a factor of 2.7 \cite{CRADPropEL15}, compared to that measured at room temperature \cite{ArELExp08} and to that expected from the theory \cite{ArELTheory11}. This enhancement was explained by a hypothesis that 50\% of emitted photons were due to N$_2$ 2PS emission in the near UV: a fraction of such an emission was assumed to be recorded directly, i.e. escaping re-emission in the WLS film and thus to have a considerably higher photon collection efficiency.

This hypothesis implies that the excitation transfer from Ar to N$_2$ species is substantially enhanced at 87 K compared to room temperature. On the other hand, the comprehensive analysis of energy levels, photon emission bands and reaction rate constants in two-phase Ar doped with N$_2$, performed recently in \cite{ArXeN2Proc17}, tends us to believe that such hypothesis can hardly be true at such a small N$_2$ content, leading to the conclusion that the experimental data of \cite{CRADPropEL15} were incorrectly interpreted. 

In this paper, we will attempt to identify the problem, in particular we will try to reproduce the experimental results of \cite{CRADPropEL15} and to understand what the mystery of proportional electroluminescence in two-phase Ar is.

The present study was performed in the course of the development of two-phase Cryogenic Avalanche Detectors (CRADs) of ultimate sensitivity for rare-event experiments \cite{CRADRev12,CRADProject12,IonYield14,CryoMPPC15,CryoPMT15,XRayYield16,CryoPMT17}.

\section{Experimental setup and measurement procedures}

\begin{figure}[ht]
	\centering
	\includegraphics[width=0.6\columnwidth,keepaspectratio]{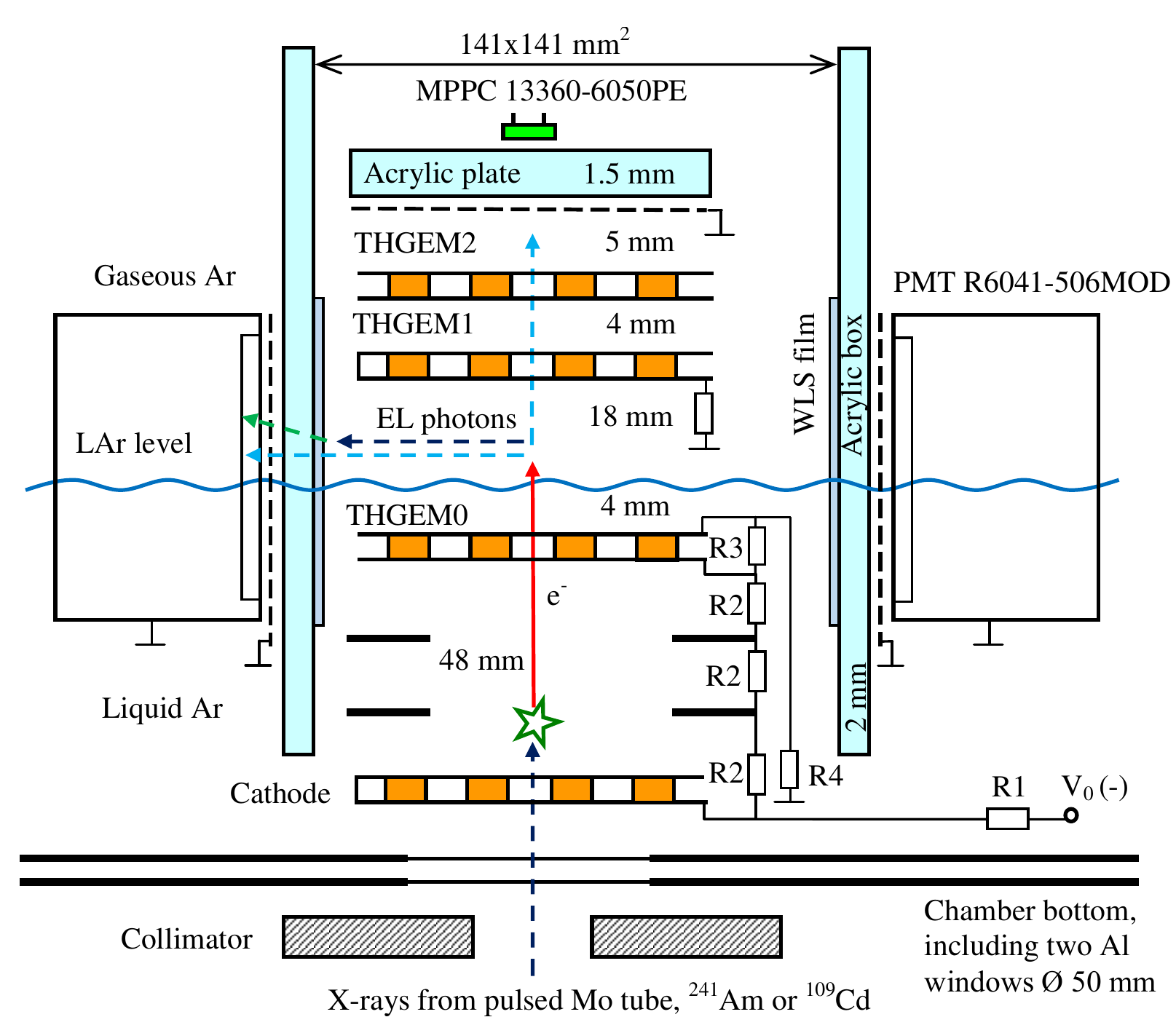}
	\caption{Schematic view of the experimental setup (not to scale). The resistors of the voltage divider have the following values: R1, R2, R3 and R4 is 80, 40, 4 and 600 M$\Omega$ respectively.}
	\label{Setup}
\end{figure}

The experimental setup of the present work was used in the measurement session of 2016; it was similar to that used in our previous measurements in 2015 \cite{CRADPropEL15,CRADELGap17}: see figure ~\ref{Setup}. The setup included a 9 l cryogenic chamber filled with 2.5 liters of liquid Ar with a minor (9 ppm) admixture of N$_2$. The detector was operated in a two-phase mode in the equilibrium state, at a saturated vapor pressure of 1.000$\pm$0.003 atm and at a temperature of 87.3 K. During each cooling procedure Ar was purified from electronegative impurities by Oxisorb filter, providing electron life-time in the liquid >100 $\mu$s \cite{CRADELGap17}. 

The cryogenic chamber comprised a cathode electrode, two field-shaping electrodes and a thick GEM (THGEM0), immersed in a 55 mm thick liquid Ar layer. These four elements were biased through a resistive high-voltage divider placed within the liquid, forming a drift region in liquid Ar, 48 mm long. A 4 mm thick liquid Ar layer above the THGEM0 acted as an electron emission region. A double-THGEM was placed in the gas phase above the liquid. The EL gap (the EL region), 18 mm thick, was formed by the liquid surface and the THGEM1 plate; the latter was grounded through a resistor acting as an anode of the gap. All electrodes had the same active area, of 10$\times$10 cm$^2$. The voltage applied to the divider varied from 11 to 22 kV, producing the electric drift field in liquid Ar of 0.34-0.68 kV/cm, electric emission field of 2.6-5.1 kV/cm and electric field in the EL gap of 4.0-8.0 kV/cm. The average drift time of the electrons across the drift, emission and EL regions varied from about 25 to 35 $\mu$s, depending on the applied electric fields. 

\begin{figure}[hbt]
	\centering
	\includegraphics[width=0.6\columnwidth,keepaspectratio]{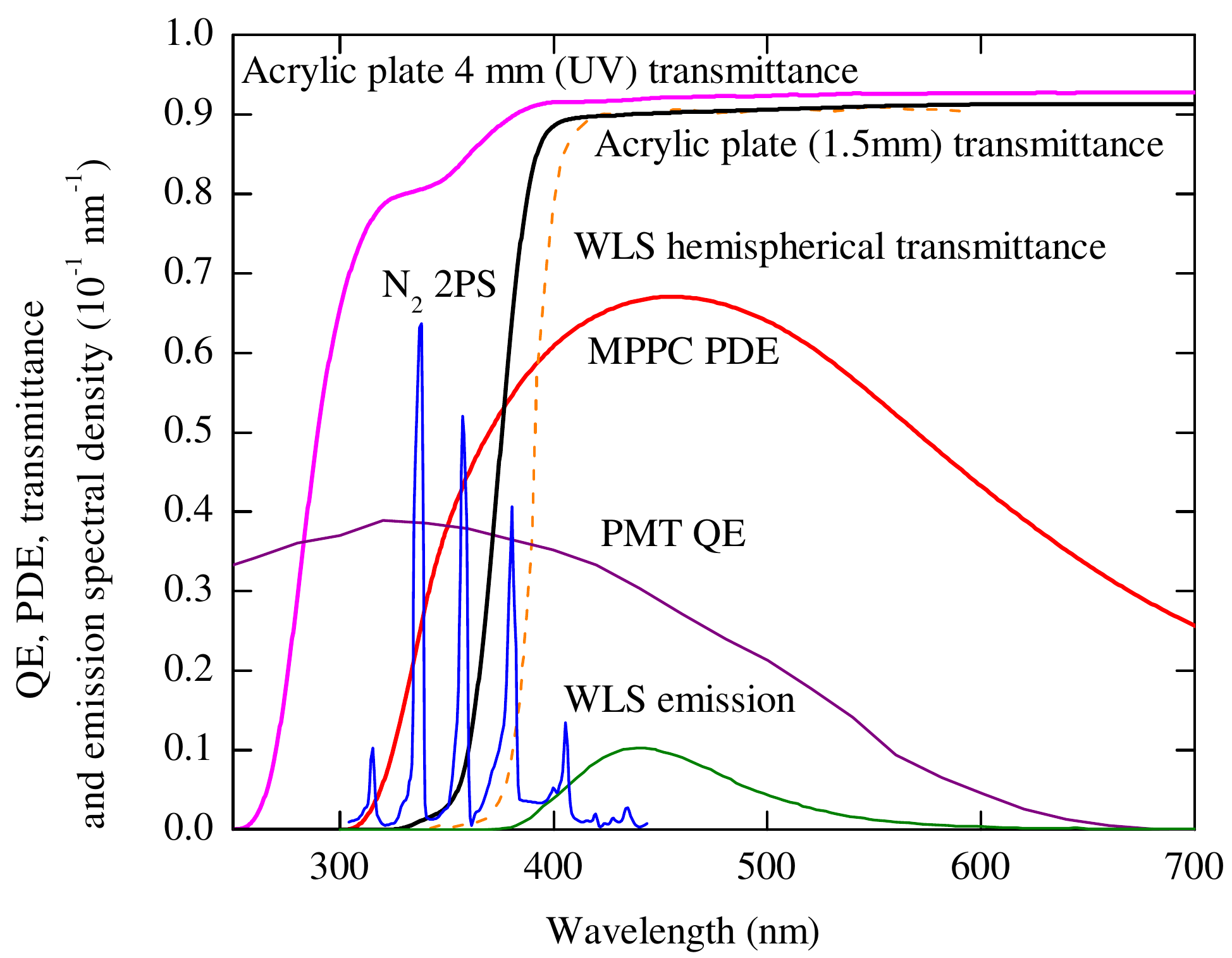}
	\caption{Quantum efficiency (QE) of the PMT R6041-506MOD at 87 K obtained from \cite{Hamamatsu,PMTQE} using a temperature dependence derived there, Photon Detection Efficiency (PDE) of the MPPC 13360-6050PE \cite{CryoMPPC15,Hamamatsu}, transmittance of the 1.5 mm thick acrylic plate in front of the MPPC measured by us  and hemispherical transmittance of the WLS (TPB in polystyrene) \cite{TPB2}. Also shown are the EL emission spectrum of Ar doped with N$_2$ (0.2\%) in the near UV and visible range measured at room temperature and high ($ \sim $1 atm) pressure \cite{Takahashi83} and that of the WLS \cite{TPB1}.}
	\label{QEPDE}
\end{figure}

The EL gap was viewed by four compact 2-inch PMTs R6041-506MOD \cite{CryoPMT15,CryoPMT17}, located on the perimeter of the gap. To prevent discharges and field penetration from the high-voltage region of the EL gap, the PMTs were electrically insulated from the gap by a grounded mesh and an acrylic protective box. To make the PMT sensitive to the VUV emission of pure Ar, four WLS films based on TPB (tetraphenyl-butadiene) in polystyrene matrix \cite{TPB2,TPB1} were deposited on the inner box surface facing the EL gap in front of each PMT. In addition, the EL gap was viewed by a multi-pixel photon counter (MPPC or SiPM) through an acrylic plate and double-THGEM assembly (the latter acting as a shadow mask), with an overall spectral sensitivity ranging from the near UV (360 nm) to the near IR (1000 nm): see figure ~\ref{QEPDE}.

To verify the reproducibility of the results, we have made two important modifications of the setup as compared to that of \cite{CRADPropEL15,CRADELGap17}: we replaced those setup elements, which might most strongly affect the accuracy of the EL yield measurements. Firstly,  we completely replaced the PMT system, including all the PMTs and the acrylic box with the WLS films. Secondly, we replaced the MPPC of \cite{CRADPropEL15,CRADELGap17} with that of 13360-6050PE type \cite{Hamamatsu}, having a 4-fold larger active area (6x6 mm$^2$) and better characteristics at cryogenic temperatures, in particular a higher gain and smaller cross-talk.

In addition to these modifications, we used a new method to measure the N$_2$ content in Ar, namely the emission-spectrum-measurement technique using the gas analyzer "SVET" \cite{SVET}, in contrast to \cite{CRADPropEL15,CRADELGap17} where the Residual-Gas-Analyzer (RGA) technique was used. Similarly to \cite{CRADPropEL15,CRADELGap17}, the N$_2$ content was measured in the gas contained in the stainless steel bottle from which the gas was liquefied into the cryogenic chamber and collected back (by cooling the bottle with liquid nitrogen),  respectively before and after the cryogenic measurement. The N$_2$ content measured this way amounted to 9$\pm$3 ppm, corresponding to the content in the two-phase mode of about 9 and 24 ppm in the liquid and gas phase respectively (according to Raoult's law). 

\begin{figure}[hbt]
	\centering
	\includegraphics[width=0.6\columnwidth,keepaspectratio]{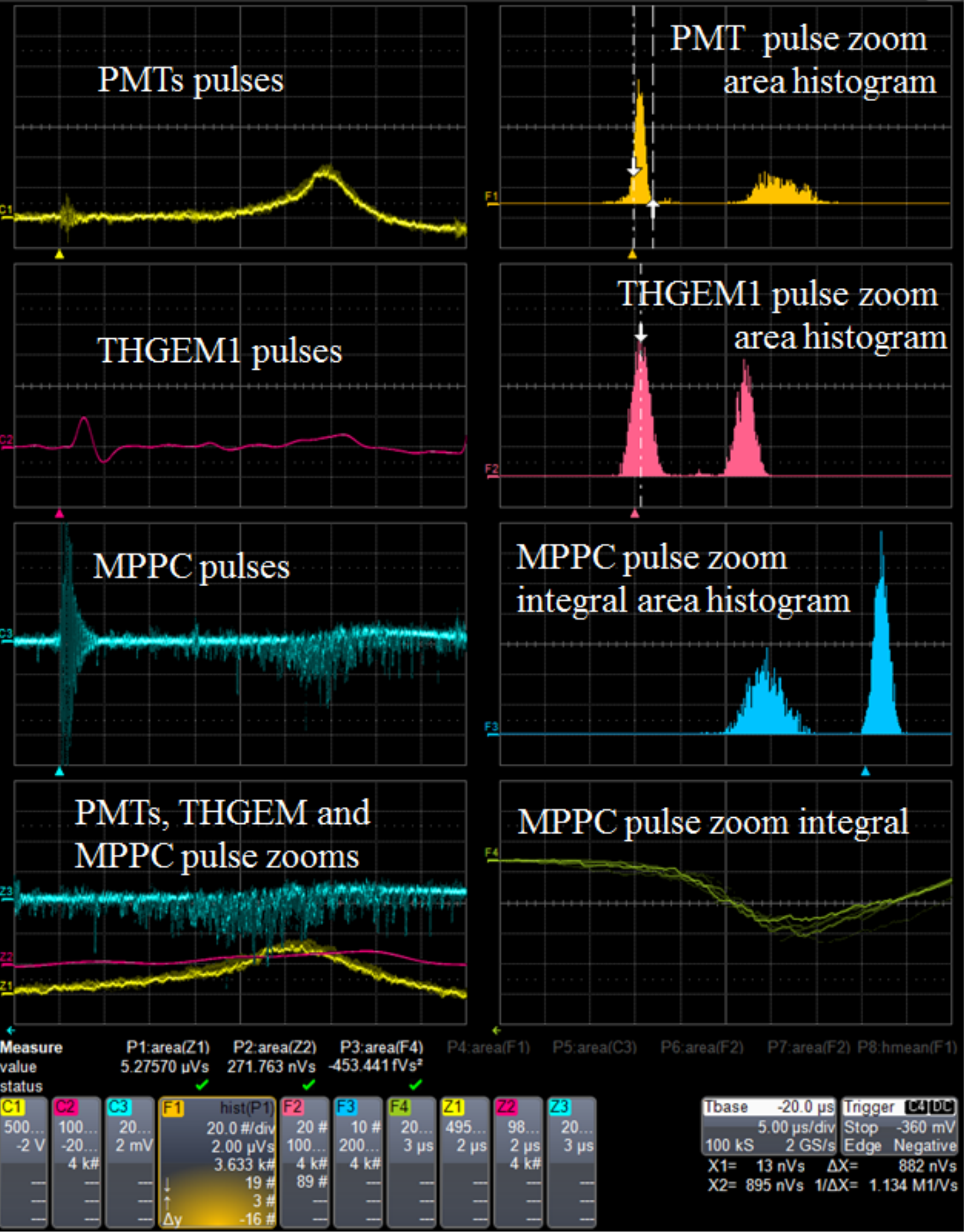}
	\caption{Typical signals from the EL gap, induced by pulsed X-rays, at an electric field of 6.84 kV/cm in the EL gap and 0.59 kV/cm in the drift region. These are shown on the left: the summed optical signal from the four PMTs, the charge signal from the gap anode (THGEM1)  and the optical signal from the MPPC. The distributions of the PMT and THGEM1 amplitudes (pulse time-integral or pulse area) and that of the MPPC (pulse double time-integral) are shown on the right, together with the electronics noise spectra.} 
	\label{Signal}
\end{figure}

The detector was irradiated from outside through a collimator and two aluminium windows (each 1 mm thick) by X-rays from a pulsed X-ray tube with Mo anode operated at a voltage of 40 kV (at a rate of 240 Hz) \cite{XRayYield16}. The X-ray pulse was strong enough, to provide a measurable ionization charge in the EL gap (tens of thousands electrons), and sufficiently fast (0.5 $\mu$s), to provide a reasonable time resolution.

Three types of signals were recorded from the EL gap: the optical signal from the PMTs, the charge signal from the THGEM1 acting as an anode of the EL gap and the optical signal from the MPPC (see figure ~\ref{Signal}). The optical signals from the four PMTs were linearly summed using CAEN N625 unit; they were amplified with a linear amplifier with a shaping time of 200 ns. The THGEM1 charge signals were recorded using charge-sensitive and shaping amplifiers with an overall time constant of 1 $\mu$s. The MPPC optical signals were recorded using a fast amplifier with a shaping time of 40 ns.

\begin{figure}[ht]
	\includegraphics[width=0.5\textwidth]{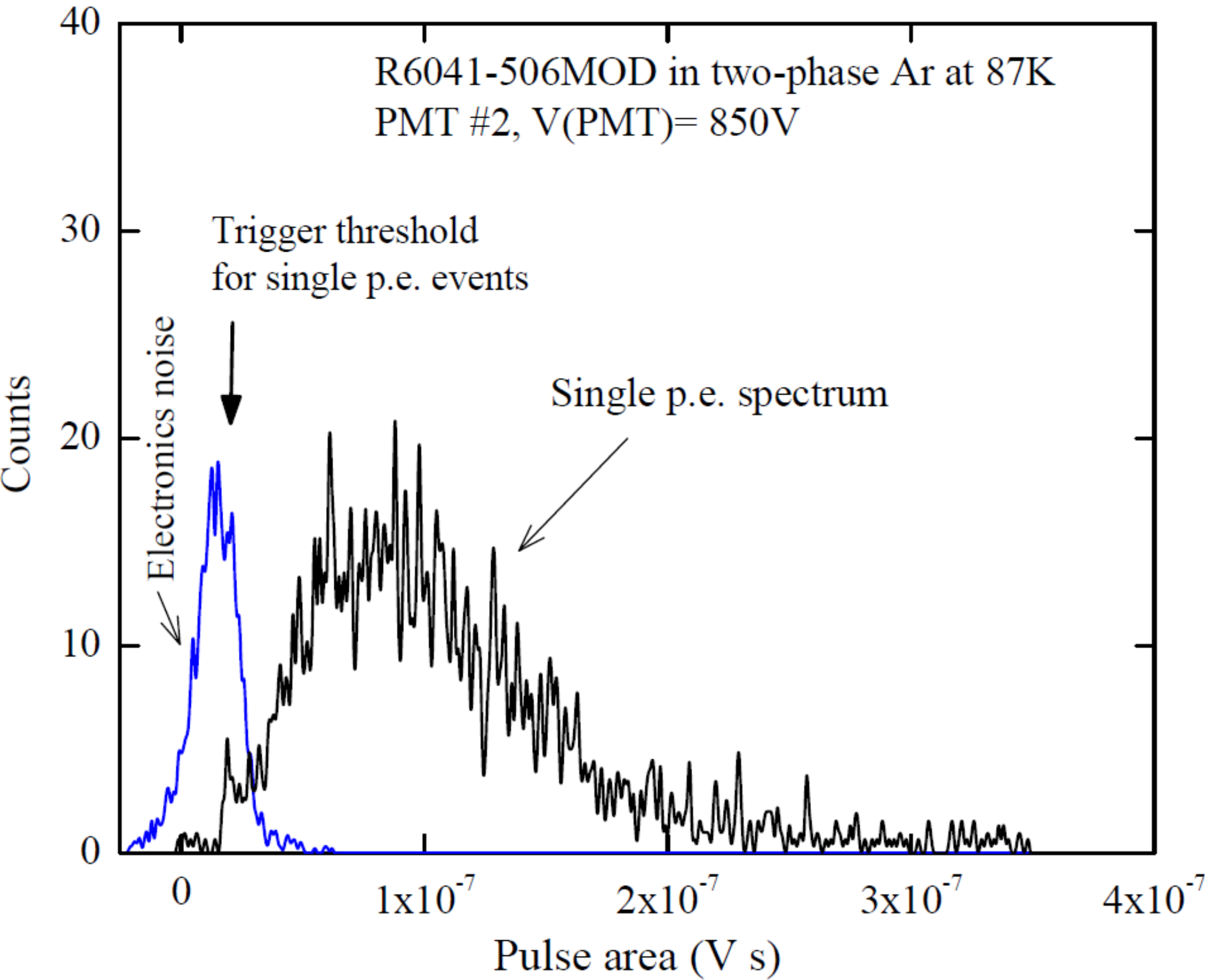}
	\hfill
	\includegraphics[width=0.5\textwidth]{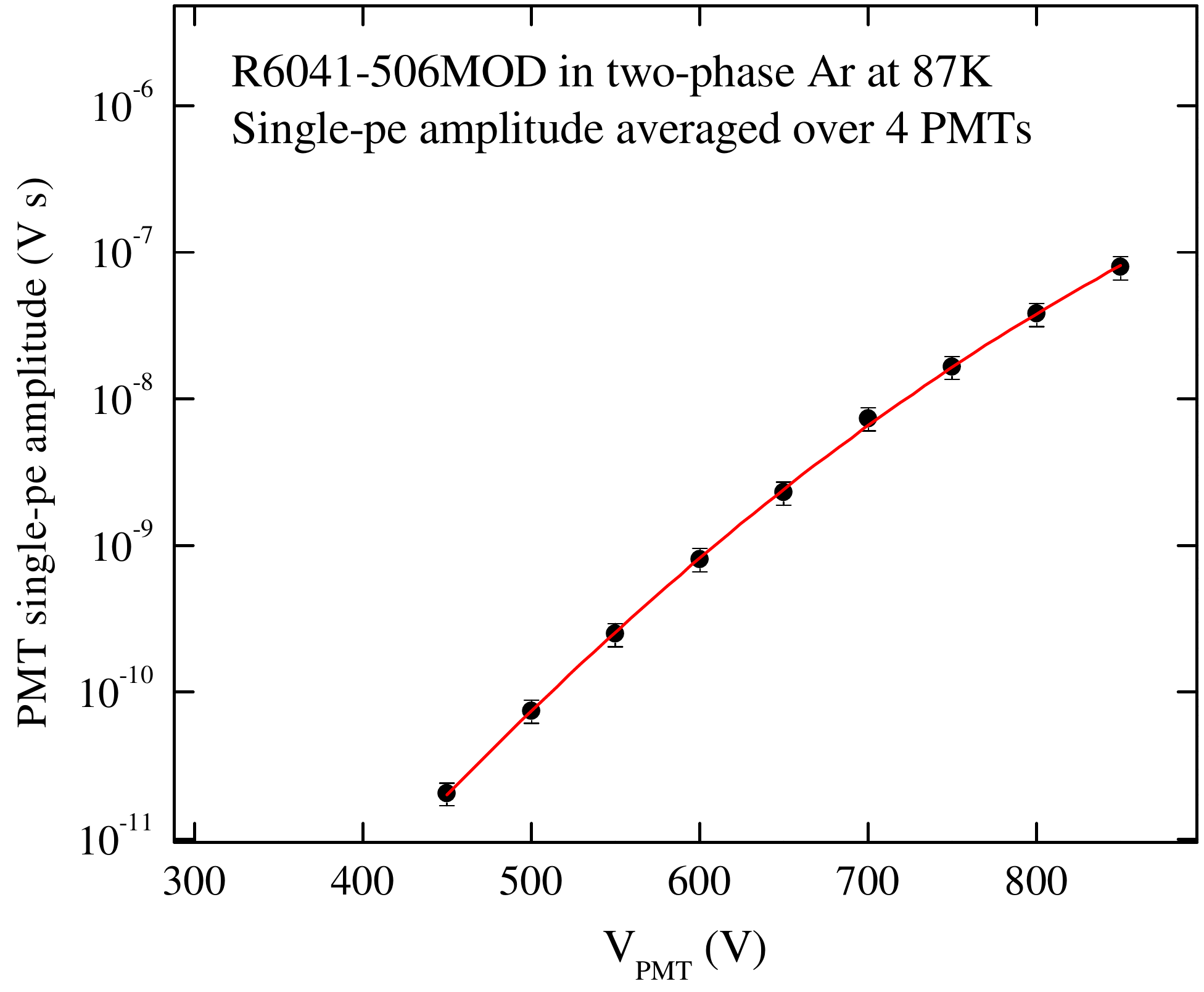}
	\caption{Left: Single-photoelectron pulse-amplitude (pulse-area) distribution of the noise signals of one of the PMTs; electronics noise spectrum is also shown. Right: PMT gain averaged over the four PMTs, in terms of the averaged PMT single p.e. amplitude, as a function of the voltage applied to the PMT divider.}
	\label{APMT}
\end{figure}

The amplitude of the PMT signal was expressed in the numbers of photoelectrons (p.e.) by measuring the area (single time-integral) of the pulse in its selected part, positive and free from electrical pickups (PMT pulse zoom in figure ~\ref{Signal}), divided by the area of the appropriate part of the single-electron pulse. The latter was determined in special measurements using PMT noise signals (figure ~\ref{APMT}, left) and PMT gain dependence on voltage (figure ~\ref{APMT}, right).

The amplitude of the THGEM1 signal was expressed in the numbers of electrons by measuring the area of the pulse in its selected part, positive and free from electrical pickups (THGEM1 pulse zoom in figure ~\ref{Signal}), using  amplifier circuit calibration, the latter procedure including the charge injection with the help of a pulse generator and a precise capacitance.

The MPPC signal was essentially bipolar, reflecting the characteristic response of the fast amplifier. Accordingly, we had to apply the double time-integration to correctly determine the signal amplitude, since the single time-integral tended to zero at given time scales (5 $\mu$s/div): see  figure ~\ref{Signal}. Similarly to \cite{THGEMGAPD10}, this technique permitted assessing the total amplitudes of long MPPC signals, consisting of multiple short single-p.e. pulses separated in time; a typical example of such signal is shown in figure ~\ref{Signal}. On the other hand, in this method the expression of the MPPC amplitude in the p.e. number turned out to be difficult, since the single p.e. amplitude at such a large time scale was not easy to determine: due fluctuations of the zero-level at large time scales, the single p.e. peaks were eroded so that they became indistinguishable. To overcome this problem, we had to measure the single p.e. amplitude at shorter times and extrapolate it to larger time scales. However this procedure  was not that accurate. In this work we applied a new improved procedure to associate the MPPC signal amplitude with the p.e. number. In this procedure, performed off-line, we directly counted the number of p.e. peaks in the pulse waveforms on a limited data set and related it to the measured amplitude.

It should be remarked that the MPPC single p.e. amplitude in our previous work \cite{CRADPropEL15} was measured at much shorter integration times than those used in the EL yield measurements, and thus was essentially underestimated. This resulted in the overestimated MPPC amplitude, which in turn lead to the overestimated contribution of the N$_2$ 2PS emission component to the overall EL yield, as will be discussed in the next section. 

The sensitivity of the two-phase detector with EL gap is characterized by the EL gap yield. In our case it can be defined as the number of photoelectrons recorded by the PMTs or MPPC ($N_{pe}$), per drifting electron in the EL gap: $Y_{gap}=N_{pe}/N_e $. Here $N_e$ is the charge expressed in the number of electrons, drifting in the EL gap and producing proportional electroluminescence.

Figure ~\ref{QEPDE} presents optical spectra of the PMT Quantum Efficiency (QE) \cite{Hamamatsu,PMTQE}, MPPC Photon Detection Efficiency (PDE) \cite{CryoMPPC15,Hamamatsu}, acrylic plate transmittance (measured by us) and WLS (TPB in polystyrene) hemispherical transmittance \cite{TPB2}. In addition, the EL emission spectrum of Ar doped with N$_2$, namely the 2PS spectrum \cite{Takahashi83}, and the emission spectrum of the WLS (TPB in polystyrene) \cite{TPB1} are presented. These data were used to determine the absolute EL yield. 

The absolute EL yield is defined as the number of emitted photons ($N_{ph}$) normalized to the number of drifting electrons producing electroluminescence and to the electron drift path ($d$):
\begin{equation}
\label{eq.1} Y_{EL}=N_{ph}/N_e/d \,.
\end{equation}
The number of photons recorded by the PMTs or MPPC is defined as
\begin{equation}
\label{eq.1} N_{ph}=N_{pe}/PCE \,.
\end{equation}
Here $PCE$ is the photon-to-photoelectron conversion efficiency. For PMTs, it is equal to $PCE=\varepsilon <CE> <QE>$ for the VUV and UV (N$_2$ 2PS) components re-emitted by the WLS and $PCE=\varepsilon <QE>$ for the UV component recorded directly.   For MPPC, it is equal to $PCE=\varepsilon <PDE>$. 

Here $\varepsilon$ is the photon collection efficiency, $<CE>$ the WLS conversion efficiency averaged over the VUV (pure Ar) or UV (N$_2$ 2PS) emission spectrum, $<QE>$ and $<PDE>$ the PMT QE and MPPC PDE averaged over the WLS or N$_2$ 2PS emission spectrum and appropriately convoluted with the WLS hemispherical or acrylic plate transmittance spectrum  (figure ~\ref{QEPDE}). In particular, PMT QE averaged over the WLS emission spectrum is $<QE>=26.6\%$, PMT QE averaged over the N$_2$ 2PS emission spectrum and convoluted with the WLS hemispherical transmittance is $<QE>=5.1\%$ and MPPC PDE averaged over the N$_2$ 2PS emission spectrum and convoluted with the acrylic plate transmittance is $<PDE>=18.1\%$. Also,  we used $CE=0.58$  around 128 nm \cite{TPB1}, $CE=0.40$ at 300-400 nm \cite{TPB1,TPB2,TPB3} and $CE=0$ above 400 nm.

The photon collection efficiencies were calculated using several Monte-Carlo simulation procedures. The first and the second ones simulated the emission of the photons in the EL gap of either the VUV (128 nm) or the UV (N$_2$ 2PS) component absorbed in the WLS. The latter amounted to 85\% of the N$_2$ 2PS, at 300-400 nm, obtained from convolution of the N$_2$ 2PS emission and WLS hemispherical transmittance spectra of figure ~\ref{QEPDE}. The propagation of photons to the WLS film, their conversion there to the visible light photons and their further propagation to the PMT photocathodes were simulated. 

The third procedure simulated the emission of the photons of the UV (N$_2$ 2PS) component escaping absorption in the WLS (amounting to 15\% of the N$_2$ 2PS, at 400-450 nm) and thus directly propagating to the PMTs. The photon-to-photoelectron conversion efficiency for this component turned out to be considerably higher than that of the light re-emitted by the WLS, by about a factor of 20, due to the absence of re-emission and total internal reflection losses.

The fourth procedure simulated the emission of the photons of the UV (N$_2$ 2PS) component and their propagation to the MPPC through the double-THGEM, protective grid and acrylic plate (amounting to 31\% of the N$_2$ SPS, at 360-450 nm).

Because of the critical dependence of the absolute photon yield on given simulation procedures, the latter were carried out in two ways using different program codes: that of developed ourselves and that using photon tracing facility of Geant4 software package. The two ways were consistent with an uncertainty of 20\%, which was finally included in systematic error. 

The results of these simulation procedures are presented in table ~\ref{tab.1} showing the values of $\varepsilon$, <QE>, <PDE> and PCE for appropriate emission components. 

\begin{table}
	\caption{\label{tab.1} Photon collection efficiency ($\varepsilon$) calculated for different photon emission components, WLS conversion efficiency averaged over the VUV (pure Ar) or UV (N$_2$ 2PS) emission spectrum (<CE>), quantum efficiency of PMTs (<QE>) or photon detection efficiency of MPPC (<PDE>) averaged over the WLS or N$_2$ 2PS emission spectrum and appropriately convoluted with the WLS hemispherical or acrylic plate transmittance spectrum, and the resulting photon-to-photoelectron conversion efficiency (PCE). }
	\begin{center}
		\begin{tabular}{|cccccc|}
			\hline
			Emission component & Photodetector & $\varepsilon$ & <CE> & <QE> or <PDE> & PCE \\
			\hline
			VUV (pure Ar) at 128 nm & 4PMT & 0.0052 & 0.58 & 0.266  & 8.0$\times$10$^{-4}$ \\
			re-emitted by WLS & & & & & \\
			UV (N$_2$ 2PS) at 300-450 nm & 4PMT & 0.0061 & 0.34 & 0.266 & 5.5$\times$10$^{-4}$ \\
			re-emitted by WLS & & & & & \\
			UV (N$_2$ 2PS) at 300-450 nm & 4PMT & 0.052 & - & 0.051  & 2.7$\times$10$^{-3}$ \\
			recorded directly & & & & & \\
			UV (N$_2$ 2PS) at 300-450 nm & MPPC & 0.00032 & - & 0.181  & 5.9$\times$10$^{-5}$ \\
			recorded directly & & & & & \\				
			\hline
		\end{tabular}
	\end{center}
\end{table}

Other details of the experimental setup and measurement procedures were described elsewhere \cite{CRADPropEL15,CRADELGap17}.

\section{Experimental results}

In the previous paper \cite{CRADPropEL15}, it was two observations that prompted us to propose a hypothesis about the enhancement of the N$_2$ 2PS emission in two-phase Ar. The first observation is related to the excess of the EL yield compared to that expected from the theory, by a factor of 2.7, within the "pure Ar" approach, i.e. when the photon emission is fully attributed to that of Ar$^{\ast}_2$ excimers in the VUV. The second observation is related to the presence of the MPPC signal, directly indicating on the existence of emission component in the non-VUV (see figure ~\ref{QEPDE} for spectra reference).

\begin{figure}[hbt]
	\centering
	\includegraphics[width=0.6\columnwidth,keepaspectratio]{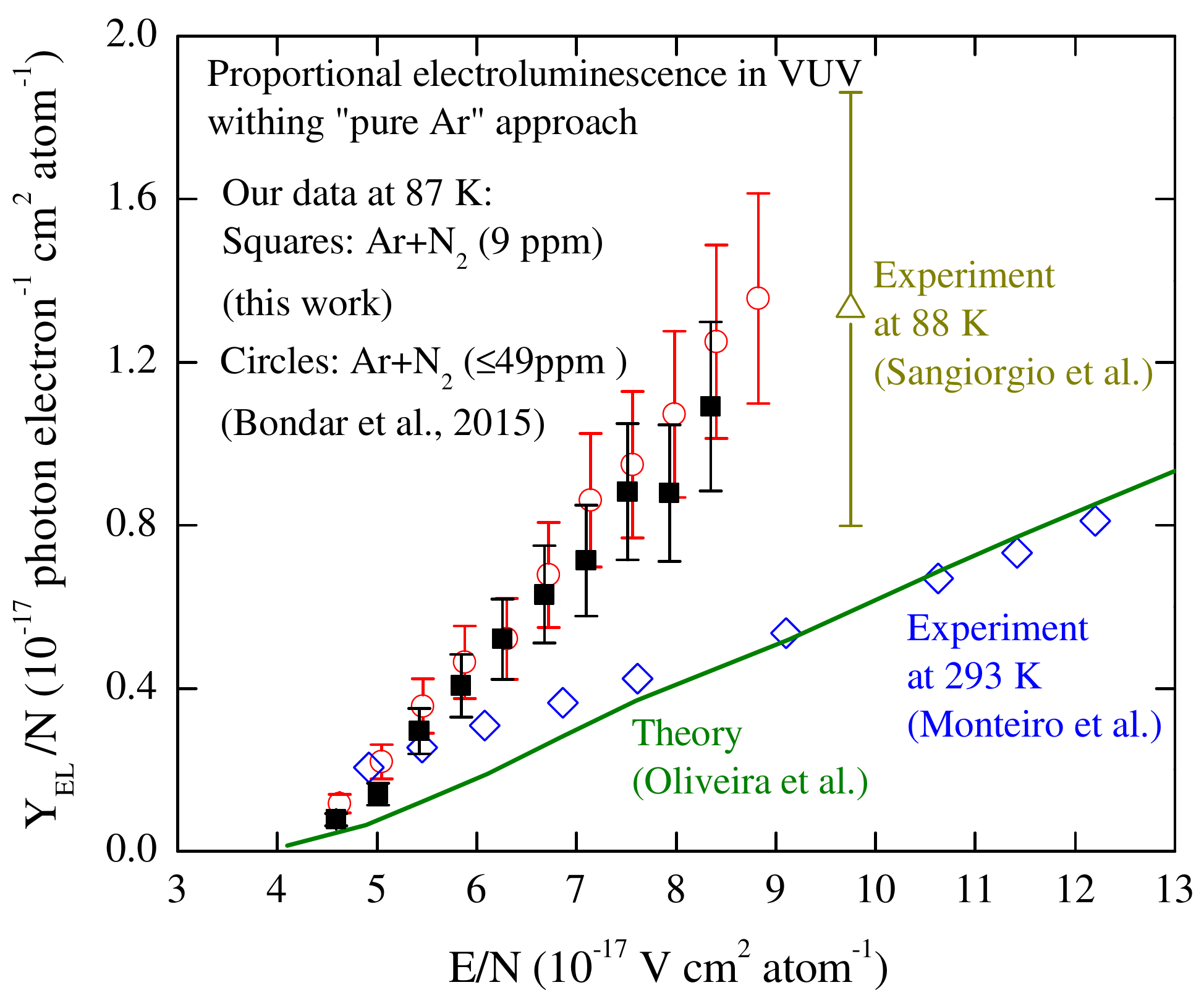}
	\caption{Reduced EL yield in the VUV as a function of the reduced electric field, obtained in two-phase Ar within the "pure Ar" approach. Shown are the data of the present work  obtained in 2016 in Ar+N$_2$~(9 ppm), those of our previous work obtained in 2015 in Ar+N$_2$~($\leq$49 ppm) (Bondar et al. \cite{CRADPropEL15}) and the data point deduced from the work of Sangiorgio et al. \cite{Sangiorgio13} obtained in pure Ar. For  comparison, the yields in gaseous Ar in the VUV obtained experimentally at 293 K (Monteiro et al. \cite{ArELExp08}) and theoretically (Oliveira et al. \cite{ArELTheory11}) are presented.}
	\label{PEPureAr}
\end{figure}

The first observation, i.e. the "pure Ar" approach, is illustrated in figure ~\ref{PEPureAr}, showing the reduced EL yield (Y$_{EL}$/N) as a function of the reduced electric field (E/N). On the one hand, one can see that the data of the present work strongly support the result of our previous work that the EL yield in two-phase Ar is substantially enhanced (by a factor of 2-3) compared to that predicted by the theory and to that  measured at room temperature. On the other hand, there is a significant difference in N$_2$ content between the present data (of 2016) and the previous data (of 2015): 9 ppm versus 49 ppm. This fact indicates that the enhancement effect might not be related to N$_2$ dopant, in contrast to that assumed in \cite{CRADPropEL15}. These conclusions are also supported by the data point of another group obtained for pure Ar, deduced by us from the work \cite{Sangiorgio13} and shown in figure ~\ref{PEPureAr}. 

\begin{figure}[hbt]
	\centering
	\includegraphics[width=0.6\columnwidth,keepaspectratio]{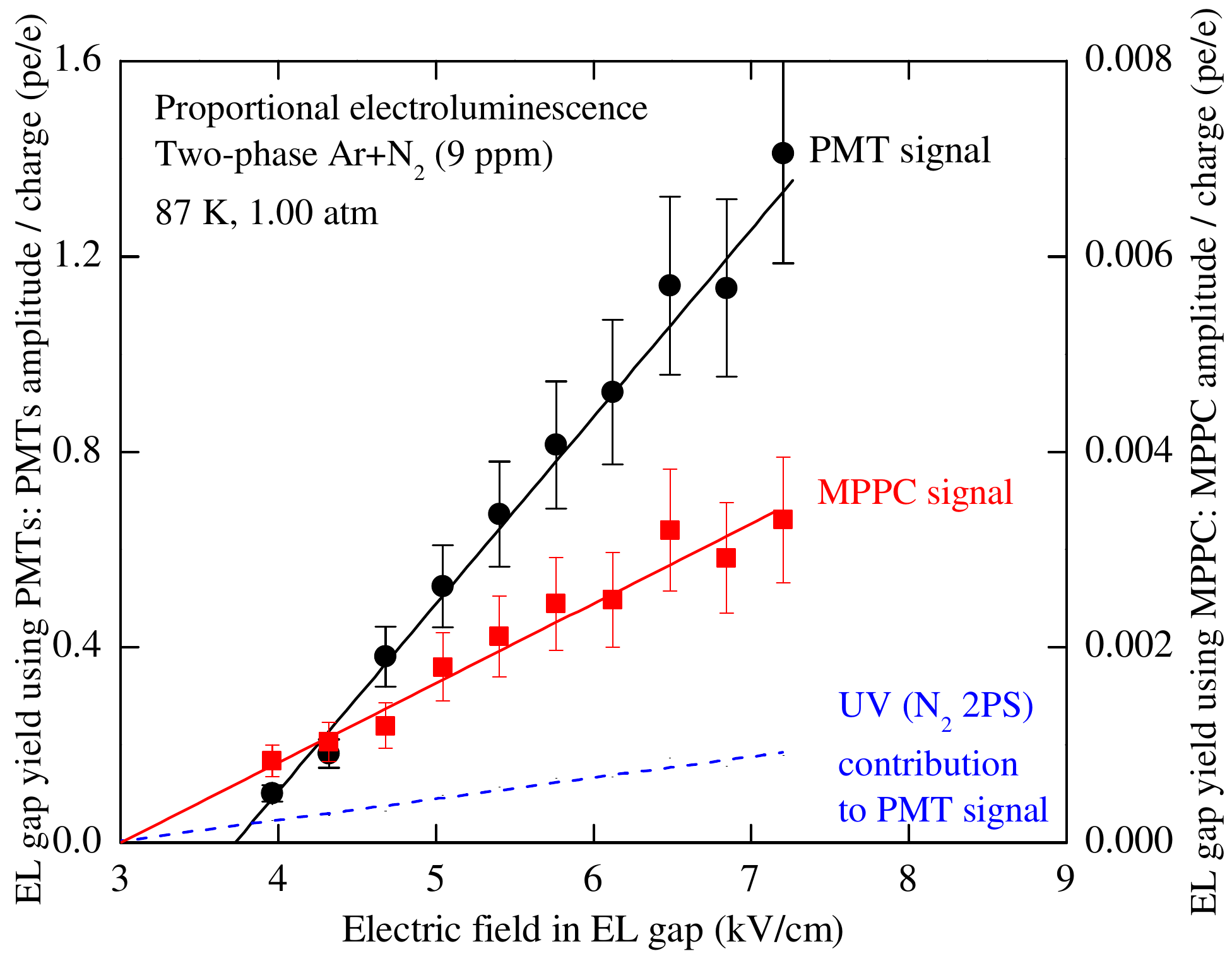}
	\caption{EL gap yield measured using PMT signal (left scale) and MPPC signal (right scale) as a function of the electric field in the EL gap. The dashed line indicates the contribution of proportional electroluminescence in the UV within the "Ar+N$_2$" approach (i.e. due to N$_2$ 2PS emission) to the PMT signal, deduced from the EL yield using MPPC signal.
	}
	\label{PEUV}
\end{figure}

The second observation is illustrated in figure ~\ref{PEUV}, showing the EL gap yield measured using PMT and MPPC signals. On the one hand, we confirm here the presence of MPPC signal observed in our previous work \cite{CRADPropEL15}, consequently confirming the existence of non-VUV emission. To move forward, we have to adopt a certain model for this non-VUV component. An exhaustive description of sources of non-VUV emission in two-phase Ar doped with N$_2$ and Xe can be found elsewhere \cite{ArXeN2Proc17}, presenting comprehensive analysis of energy levels, photon emission bands and reaction rate constants. In the PMT sensitivity region, all such sources are related to N$_2$ emission, namely to that of 2PS or 1PS (first positive system) due to N$_2^{\ast}(C)$ or N$_2^{\ast}(B)$ excited states respectively, the first source being preferred according to reaction kinetic analysis. Also, emission in the NIR (near infrared) can not be related to this non-VUV component, since NIR electroluminescence starts at much higher fields \cite{CRADProject12}. Given no other rational assumptions, we adopt here the hypothesis of N$_2$ 2PS origin of non-VUV emission, similarly to \cite{CRADPropEL15}.

%Similarly to \cite{CRADPropEL15}, it can be associated with N$_2$ 2PS emission in the near UV, since we are not aware of any other potential sources of non-VUV emission. 

On the other hand, when we estimated the contribution of this UV (N$_2$ 2PS) component to the PMT signal, it turned out to be a factor of 6 smaller than that reported in \cite{CRADPropEL15}: see figure ~\ref{PEUV}. We think that this discrepancy is due to the fact that the MPPC single p.e. amplitude in our previous work \cite{CRADPropEL15}  was essentially underestimated, leading to overestimated MPPC amplitude, as discussed in the previous section. Accordingly, the UV contribution amounts to 14\% only, the rest of the PMT signal being attributed to VUV component. 

\begin{figure}[hbt]
	\centering
	\includegraphics[width=0.6\columnwidth,keepaspectratio]{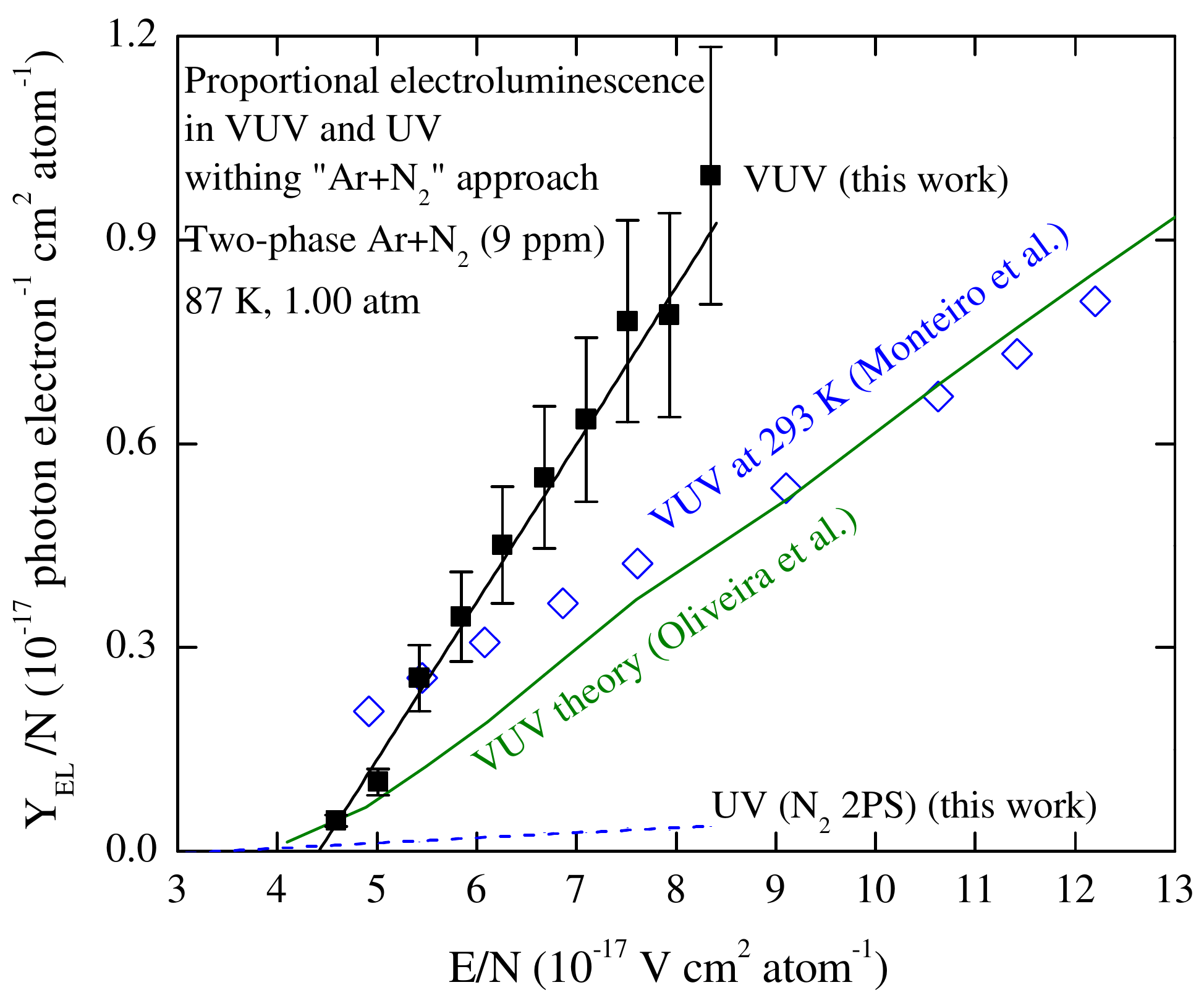}
	\caption{
		Reduced EL yield in the VUV and UV as a function of the reduced electric field, obtained in this work within the "Ar+N$_2$" approach. For  comparison, the yields in gaseous Ar in the VUV obtained experimentally at 293 K (Monteiro et al. \cite{ArELExp08}) and theoretically (Oliveira et al. \cite{ArELTheory11}) are presented.
	}
	\label{PE2PS}
\end{figure}

Figure ~\ref{PE2PS} shows the reduced EL yield in two-phase Ar, obtained within this "Ar+N$_2$" approach, i.e. in the VUV and UV assuming that the latter component (recorded via MPPC) is due to N$_2$ 2PS emission. In this approach, the discrepancy with the theory for the VUV emission still remains quite significant, of a factor of 2. 

The linear dependence of the EL yield in the VUV (Ar$^{\ast}_2$ excimer emission) and in the UV (N$_2^{\ast}(C)$ 2PS emission) on the electric field (figure ~\ref{PE2PS}) is described by the following equations:
\begin{equation}
Y_{EL}/N = 0.231E/N-1.024 \,,
\end{equation}
\begin{equation}
Y_{EL}/N = 0.008E/N-0.026 \,.
\end{equation}
The slope of the line defines the amplification parameter of proportional electroluminescence. It amounts to 231$\pm$13 and 8$\pm$1 photons per kV in the VUV and UV respectively. 

We may conclude, that the contribution of UV component is insufficient to explain the enhancement effect of the PMT signal, in contrast to \cite{CRADPropEL15} and in accordance with theoretical analysis of \cite{ArXeN2Proc17}.
As a result, the situation with interpretation of data becomes even more confusing than it was in the previous work \cite{CRADPropEL15}. We still remain with the following two questions for proportional electroluminescence in two-phase Ar which we cannot explain at the moment: the effect of EL yield enhancement in the VUV and the origin of non-VUV emission. 

\section{Conclusions}

In this paper, we continued to study the problem of proportional electroluminescence in two-phase Ar doped with N$_2$, with the intention of verifying the data of our previous work \cite{CRADPropEL15}. On the one hand, the results of present work confirm the effect of EL yield enhancement in two-phase Ar observed in \cite{CRADPropEL15}. We also confirm the presence of non-VUV component in addition to that of VUV, in proportional electroluminescence. On the other hand, the contribution of UV component determined here within the model of N$_2$ 2PS emission, turned out to be insufficient to explain the enhancement effect, in contrast to \cite{CRADPropEL15} and in accordance with theoretical analysis of \cite{ArXeN2Proc17}.

Finally, the present study helped us to identify the problem: at the moment we cannot explain the mystery of proportional electroluminescence in two-phase Ar; it is to be resolved in future studies.

This study was supported by Russian Science Foundation (project No. 16-12-10037); it was done within the R\&D program for the DarkSide-20k experiment. 


\end{document}